\documentclass{aastex}
\usepackage{emulateapj5,graphicx}
\begin{document}

\newcommand{\bm}[1]{\mbox{\boldmath $#1$}}
\title{
Secondary Star Formation in a Population III Object
}
\author{Hajime Susa\altaffilmark{1} 
\vskip 0.2cm
\affil{Department of Physics, Rikkyo University, Nishi-Ikebukuro,
Toshimaku, Japan}
\vskip 0.3cm
Masayuki Umemura\altaffilmark{2}
\vskip 0.2cm
\affil{Center for Computational Sciences, University of
  Tsukuba, Japan}
\altaffiltext{1}{susa@rikkyo.ac.jp}
\altaffiltext{2}{umemura@rccp.tsukuba.ac.jp}
}

\begin{abstract}
We explore the possibility of subsequent star formation 
after a first star forms in a Pop III object, 
by focusing on the radiation hydrodynamic (RHD) feedback
brought by ionizing photons as well as H$_2$ dissociating photons.
For the purpose, we perform three-dimensional RHD
simulations, where the radiative transfer of ionizing photons
and H$_2$ dissociating photons from a first star
is self-consistently coupled with hydrodynamics based on
a smoothed particle hydrodynamics method.
As a result, it is shown that density peaks above a threshold density
can keep collapsing owing to the shielding of H$_2$ dissociating radiation
by an H$_2$ shell formed ahead of a D-type ionization front. 
But, below the threshold density, an M-type ionization front 
accompanied by a shock propagates, and density peaks are radiation 
hydrodynamically evaporated by the shock. 
The threshold density is dependent on the distance from a source star, 
which is $\approx 10^2 {\rm cm^{-3}}$ for the source distance of 30pc.
Taking into consideration that the extent of a Pop III object is $\approx 100$pc
and density peaks within it have the density of $10^{2-4}$cm$^{-3}$, 
it is concluded that the secondary star formation is allowed 
in the broad regions in a Pop III object. 

\end{abstract}


\section{Introduction}
\label{intro}
In the last decade, the formation of Population III (hereafter Pop III) stars 
has been explored extensively
(Bromm, Coppi \& Larson 1999, 2002; 
Nakamura \& Umemura 1999, 2001; Abel, Bryan, Norman 2000, 2002;
Yoshida 2006).
The Pop III stars can significantly 
affect the reionization history in the universe 
\citep{Cen03,Cia03,Wyi04,Somer03,Sokasian04,Murakami05},
and the metal enrichment of intergalactic medium 
\citep{NU01,Scan02,Ricotti04}. 
In previous analyses 
\citep{Tegmark97,Fuller00,Yoshida03},
it is shown that Pop III objects 
have the halo mass of $\approx 10^6 M_\odot$ and 
the extent of $\approx 100$pc. In Pop III objects, density peaks
collapse owing to H$_2$ cooling, forming cloud cores with
the density of $10^{2-4}$cm$^{-3}$ \citep{Bromm02}.
A highest peak in the halo collapses earlier to form a first Pop III star.
Hence, subsequently collapsing cores are affected by
the radiative feedback by the first star. 
However, the range of the feedback and the possibility of 
secondary star formation in a Pop III object are still under debate. 

If a first star is distant by more than 1pc,
dense cores are readily self-shielded from the ultraviolet (UV) radiation
\citep{TU98, Kitayama01,SU04a,SU04b,Kitayama04,Dijkstra04,Alvarez06}. 
Thus, the photoevaporation
by UV heating is unlikely to work devastatingly. 
However, the photodissociating radiation (11.18-13.6eV) 
of hydrogen molecules (H$_2$) in Lyman-Werner band (LW band)
can preclude the core from collapsing, 
since H$_2$ are the dominant coolant to enable the core collapse
\citep{Haiman97,ON99,Haiman00,GB01,Machacek01}.
Hence, this may lead to momentous negative feedback.

On the other hand, ionizing radiation 
for hydrogen ($\geq$13.6eV) drives an ionization
front (I-front), which propagates in a collapsing core. 
The enhanced fraction of electrons promotes H$_2$ formation 
\citep{SK87, KS92, SUNY98, OhH02}. In particular,
the mild ionization ahead of the I-front can generate an H$_2$ shell,
which potentially shields H$_2$ dissociating photons \citep{Ricotti01}.
This mechanism is likely to work positively to form Pop III stars. 
In practice, the propagation of I-front is complex. 
When UV irradiates a dense core, the I-front changes from R-type 
on the surface to D-type inside the core. The transition occurs
via an intermediate type (M-type), which is accompanied with
the generation of shock \citep{Kahn54}. The shock can affect
significantly the collapse of the core. This is a totally radiation
hydrodynamic (RHD) process, which has not been hitherto explored 
in detail as feedback by first stars. 

In this Letter, we scrutinize the radiation hydrodynamic (RHD) feedback 
by a first star through the propagation of I-front in a dense core. 
For the purpose, we solve the three-dimensional RHD,
where the radiative transfer of ionizing photons
as well as H$_2$ dissociating photons is self-consistently coupled
with hydrodynamics. 
In \S 2, the key physical processes associated with I-front 
propagation are overviewed. The setup of simulation is
described in \S 3, and numerical results are presented in \S 4.
\S 5 is devoted to the conclusions. 


\section{Basic Physics}

If a collapsing core is irradiated by an ionizing source located
at the distance $D$, the propagation speed of I-front 
in the core is given by
\begin{equation}
v_{\rm IF}=
21~ {\rm km~s^{-1}}
\left(\frac{\dot{N}_{\rm ion}}{10^{50}{\rm s^{-1}}}\right)
\left(\frac{D}{20{\rm pc}}\right)^{-2}
\left(\frac{n_{\rm core}}{10^3{\rm cm^{-3}}}\right)^{-1},
\end{equation}
where $\dot{N}_{\rm ion}$ is the number of ionizing photons per
unit time and $n_{\rm core}$ is the number density in the cloud core. 
The sound speed in the regions cooling by H$_2$ 
is $a_1 \approx 1{\rm km~s^{-1}}$, 
while that in the ionized regions is $a_2 \approx 10{\rm km~s^{-1}}$. 
Hence, $v_{\rm IF} > 2a_2$ for low density cores or 
a nearby ionizing source, and therefore I-front
becomes R-type. 
Supposing the cloud core size $r_{\rm core}$ is on the order of 
$a_1\sqrt{\pi/G\rho_{\rm core}}$, 
the propagation time of R-type front across the core satisfies
\begin{eqnarray}
t_{\rm IF}\equiv \frac{r_{\rm core}}{v_{\rm IF}} <
 \frac{a_1}{2a_2}
\sqrt{\frac{\pi}{G\rho_{\rm core}}} 
< \sqrt{\frac{\pi}{G\rho_{\rm core}}}.
\end{eqnarray}
This means that an R-type front sweeps the core before the core 
collapses in a free-fall time. Thus, the core is likely to undergo
photoevaporation. 
On the other hand, if the density of cloud core
is high enough and the source distance is large, then
$v_{\rm IF} < a_1^2/2a_2$ and a D-type I-front emerges. 
The propagation time of D-type front across the core is 
\begin{eqnarray}
t_{\rm IF}  >
 \frac{2a_2}{a_1}\sqrt{\frac{\pi}
{G\rho_{\rm core}}} > \sqrt{\frac{\pi}{G\rho_{\rm core}}}.
\end{eqnarray}
Thus, the core can collapse 
before the I-front sweeps the core. 

The above arguments are based on the assumption that
the ionizing photon flux does not change during the propagation
of I-front.
However, the core could be self-shielded from the ionizing
radiation if 
$\dot{N}(\pi r_{\rm core}^2/4\pi D^2)<4\pi r_{\rm core}^3 
n_{\rm core}^2 \alpha_{\rm B}/3$, where $\alpha_{\rm B}$
is the recombination coefficient to all excited levels of hydrogen.
The critical density for self-shielding is given by
\begin{eqnarray}
&\;& n_{\rm shield} \simeq \left(\frac{3\dot{N}_{\rm ion}}
{16\pi D^2 a_1 \alpha_{\rm B}} \sqrt{\frac{G m_{\rm p}
}{\pi}}\right)^{2/3}\nonumber\\
&=& 5.1~ {\rm cm^{-3}}
\left(\frac{\dot{N}_{\rm ion}}
{10^{50}{\rm s^{-1}}}\right)^{2/3}\left(\frac{D}
{20{\rm pc}}\right)^{-4/3}
\left(\frac{a_1}{1{\rm km~ s^{-1}}}\right)^{-2/3}. \label{eq:shield}
\end{eqnarray}
If $n_{\rm core}> n_{\rm shield}$, the ionizing photon flux 
diminishes significantly during the I-front propagation.
Hence, even if the I-front is R-type on the surface of
cloud core, the front can change to M-type accompanied with shock,
and eventually to D-type inside the core (Kahn 1954). 

In comparison to ionizing photons, H$_2$ dissociating (LW band) 
radiation is less shielded \citep{DB96}. 
The self-shielding of LW band flux ($F_{\rm LW}$) is expressed by
\begin{equation}
F_{\rm LW} = F_{\rm LW,0} f_{\rm sh}
\left( N_{\rm H_2,14 } \right) \label{LW}
\end{equation}
where $ F_{\rm LW,0}$ is the incident flux, 
$ N_{\rm H_2,14}= N_{\rm H_2}/10^{14} {\rm cm^{-2}}$, and
\begin{equation}
f_{\rm sh}(x) = \left\{
\begin{array}{cc}
1,~~~~~~~~~~~~~~x \le 1 &\\
x^{-3/4},~~~~~~~~~x > 1 &
\end{array}
\right.
\end{equation}
However, dissociating radiation could be shielded effectively 
if an H$_2$ shell forms ahead of I-front.
This process is tightly coupled with the propagation of I-front.

\section{Setup of Simulation}
\label{simulation}
We perform radiation hydrodynamic simulations 
using a novel radiation transfer solver based 
on smoothed particle hydrodynamics (SPH)
\citep{Susa06}.
We suppose a primordial gas cloud collapsing in a run-away fashion.
The chemical compositions are initially assumed to be 
the cosmological residual value \citep{GP98}. 
The mass of cloud is $M_{\rm b} = 8.3\times 10^4 M_\odot$ 
in baryonic mass.
When the central density of cloud exceeds a certain value, 
$n_{\rm on}$, we ignite a 120$M_\odot$ Pop III star 
($\dot{N}_{\rm ion}=1.3\times 10^{50}{\rm s^{-1}}$),
which is located at $D=20$pc for fiducial models.
We also explore the dependence on the distance of a first star
by changing the relative location between a source star and
a collapsing cloud. 
The luminosity and the effective temperature
of the source star are taken from \citet{Baraffe01}.
The fiducial models are \\
\hspace*{2mm} 
Model A --- $n_{\rm on} = 3\times 10^3 {\rm cm^{-3}}$ with H$_2$ dissociating photons but without ionizing photons \\
\hspace*{2mm} 
Model B --- $n_{\rm on} = 3\times 10^3 {\rm cm^{-3}}$ 
with ionizing and H$_2$ dissociating photons \\
\hspace*{2mm} 
Model C --- same as Model B but for $n_{\rm on} = 3\times 10^2 {\rm cm^{-3}}$ \\
The physical simulation time after the ignition of the source star is
$4$Myr except $1.54$Myr for model B, 
since the central part collapses below the resolution limit. 
In Model A, it is assumed 
that only LW band photons escape from the neighbor of a source star,
whereas ionizing photons do not because of large opacity. 
This model also can be
regarded as the reference to Model B.

\section{Results}
\subsection{Failed Collapse (Model A)}

In Fig. \ref{timeevol}, the time evolution of
density profiles along the axis of symmetry is shown for three models.
In this figure, the red curve represents the distribution at 1Myr,
at which epoch the spatial distributions of physical quantities are shown 
in Fig. \ref{3models}. 
For Model A, as shown in the left panel of Fig. \ref{timeevol}, 
the collapse of central regions stops virtually 
and forms a quasi-hydrostatic core between $\sim$1Myr and $\sim$4Myr.
The failure of collapse is caused by the photodissociation of H$_2$ by
LW band photons. 
In upper panels of Fig. \ref{3models}, the spatial distribution of H$_2$
fraction on a slice along the symmetry axis is shown 
by colored dots at the positions of SPH particles. 
In lower panels of Fig. \ref{3models}, physical quantities along 
the symmetry axis are shown at this epoch.
In Model A (left panel), the temperature is several $10^2$K in the envelope 
and $\sim 10^3$K in the central regions. 
The H$_2$ column density
is lower than $10^{14}{\rm cm^{-2}}$ in the envelope and
therefore H$_2$ is highly photodissociated by LW band radiation.
In the central regions, H$_2$ column density
exceeds $10^{14}{\rm cm^{-2}}$ and therefore
H$_2$ fraction is raised up by 
the self-shielding of LW band photons according to (\ref{LW}). 
However, this level of shielding is not enough to allow
the cloud to keep collapsing by H$_2$ cooling. 
Eventually, the collapse is halted by the thermal pressure. 
The numerical results of Model A show that gravitational 
instability is hindered by permeated H$_2$ dissociating photons. 

\subsection{H$_2$ Shielded Collapse (Model B)}
In Model B, ionizing photons are included for the same model
parameter as Model A. In this model, I-front propagates into the
cloud. Although the I-front is R-type far from the cloud center,
it changes to M-type and a shock is generated as shown by
a peak at 0.5Myr in Fig. \ref{timeevol} (middle panel).
Around 1Myr, the I-front changes to D-type, and therefore
the core collapse proceeds faster than the propagation of the I-front,
as argued in \S 2. 
The density distribution is highly changed by photoionization, 
as shown by the angelfish-shaped distribution in Fig. \ref{3models}.
In particular, it is worth noting that an H$_2$ shell forms ahead of 
the I-front due to the enhanced ionization fraction.
As a result, the H$_2$ column density is steeply raised up.
This H$_2$ shell effectively shields LW band radiation from a source star,
and the resultant H$_2$ fraction ($y_{\rm H_2}$) becomes larger by an order
of magnitude than that of Model A. 
Eventually, due to H$_2$ cooling, 
the cloud core can continue to collapse to the level
of $n_{\rm H} > 10^7 {\rm cm^{-3}}$, as shown in Fig. \ref{timeevol} (middle).
Since the only difference between Models A and B is the presence of
ionizing radiation, we can conclude that photoionization can restrain the
negative feedback effect by H$_2$ 
photodissociation. 

\subsection{Shock-driven Evaporation (Model C)}
Model C is the lower density version of Model B, where $n_{\rm on}$ 
is ten times smaller than that of Model B. 
In this case, quite similarly to Model B, the I-front becomes to M-type 
at 0.5Myr (Fig. \ref{timeevol}, right panel). 
As shown in Fig. \ref{3models},
there forms an H$_2$ shell that shields LW band radiation
from a source star. But, before the core collapses in the free-fall time,
the shock accompanied with M-type I-front sweeps up the central 
regions at $\approx$ 1Myr. 
Eventually, the shock blows out the collapsing core. 
Hence, it is concluded that if an M-type I-front passes through 
the cloud core, the whole cloud is evaporated 
radiation hydrodynamically by the shock. 

\subsection{Dependence on Position of a Source Star }
As shown above, the criterion of radiation hydrodynamic 
evaporation of a collapsing core is whether an M-type I-front 
sweeps up the core. 
By changing the position of a source star, we can obtain the
threshold density for the evaporation depending on the source distance.
As a result of numerical simulations, it is shown that 
the threshold density is $\approx 10^3 {\rm cm^{-3}}$
for $D=20$pc, $\approx 10^2 {\rm cm^{-3}}$ for $D=30$pc, 
and $\approx 10 {\rm cm^{-3}}$ for $D=50$pc. 
Thus, cloud cores with $n_{\rm core} \gtrsim 10^2 {\rm cm^{-3}}$ 
can collapse at $D \gtrsim 30$pc.
Further details of the analysis on the parameter dependence will be
described in the forth-coming full paper.

\section{Conclusions}
We have performed three-dimensional radiation hydrodynamic 
simulations to scrutinize the feedback by a first star in a Pop III
object. As a result, it has been found that a collapsing core is 
evaporated by a shock if an M-type I-front sweeps the core. 
In order for a collapsing core to evade the radiation hydrodynamic 
evaporation, the core density should exceed a threshold density 
that depends on the distance from a source star. 
Above the threshold density, the I-front is quickly changed to
the D-type, and an H$_2$ shell forms ahead of the I-front, 
which effectively shields H$_2$ dissociating radiation 
from a source star. Eventually, the core can keep collapsing
owing to H$_2$ cooling. The present numerical study has shown
that if a source star is distant by more than 30pc, then
a collapsing core denser than $\approx 10^2 {\rm cm^{-3}}$ 
is absolved from evaporation. 
Taking into account that the density of cloud cores in
a Pop III object is expected to be $10^{2-4}$cm$^{-3}$, 
stars can form in the regions of $\gtrsim 30$ pc 
even after a first star forms. 
Since the extent of a Pop III object is $\approx 100$pc,
it is concluded that 
the subsequent star formation is allowed in the broad regions 
in a Pop III object. 

\bigskip
We are grateful to N. Shibazaki for continuous encouragement, 
to T. Nakamoto and K. Ohsuga for intense discussion,
and to all the collaborators in 
{\it Cosmological Radiative Transfer Codes Comparison Project}
(astro-ph/0603199) for fruitful discussions during
the workshop in CITA and Lorentz Center in Leiden.
The analysis has been made with the {\it FIRST} simulator
at Center for Computational Sciences in University of Tsukuba and 
with computational facilities in Rikkyo University. 
This work was supported in part by Grants-in-Aid, Specially
Promoted Research 16002003 and Young Scientists (B) 17740110
from MEXT in Japan.




\setcounter{figure}{0}
\begin{figure}[ht]
\begin{center}
\includegraphics[angle=0,width=14cm]{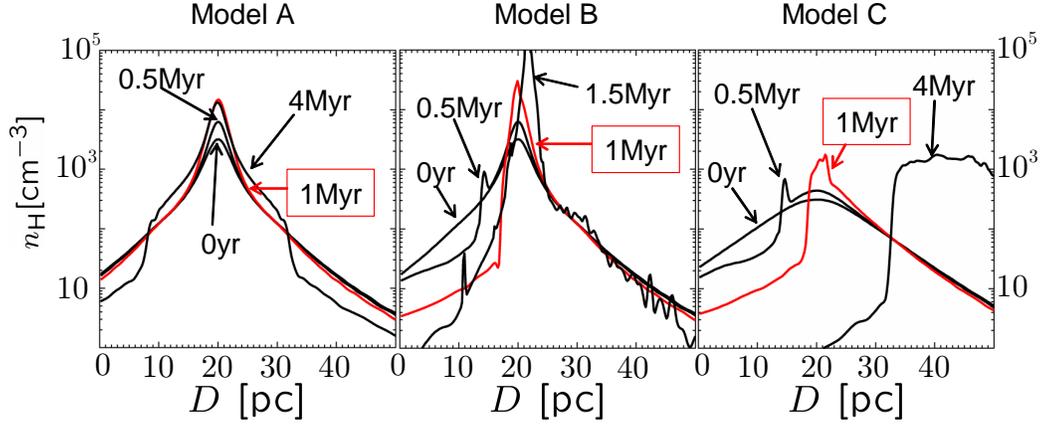}
\caption{
The time evolution of density distributions along the symmetry axis
for three models. 
Four curves correspond
to $t=0$, 0.5Myr, 1Myr, and 4Myr, for models A and C,
whereas the final plot for model B corresponds to $t=1.5{\rm Myr}$.
The red curves denote the profiles at $1{\rm Myr}$, at which
epoch the detailed structure is shown in Fig.\ref{3models}. }
\label{timeevol}
\end{center}
\end{figure}

\begin{figure}[ht]
\begin{center}
\includegraphics[angle=0,width=14cm]{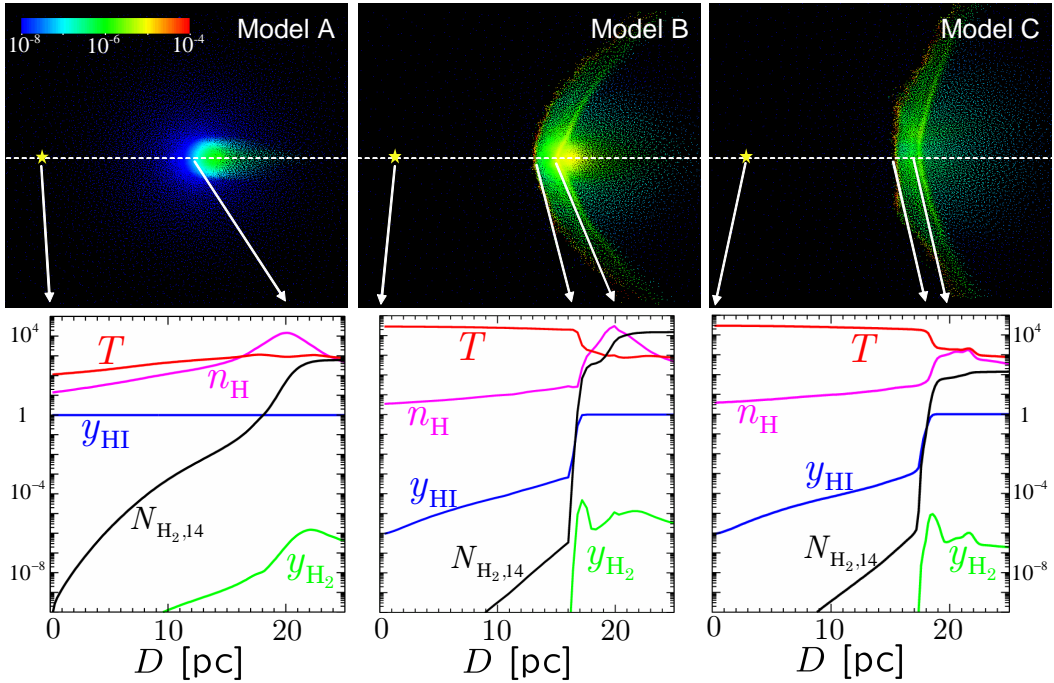}
\caption{Spatial distributions of physical quantities for three models 
at 1 Myr. Colored dots on upper three panels 
represent the distributions of SPH particles on
a slice which includes the symmetry axis (dashed line). 
The colors of particles
denote the H$_2$ fraction, whose legend is shown on the upper left. 
The yellow star represents the position of a source Pop III star. 
Lower three panels show various physical quantities along the
symmetry axis
as $T$[K], $n_{\rm H} [{\rm cm^{-3}}]$, $y_{\rm HI}$, $y_{\rm H_2}$,
and $N_{\rm H_2,14} ($H$_2$ column density from the source star in units of 
$10^{14}{\rm cm^{-2}}$).
Horizontal axis shows the distance from the source star. 
Arrows show the correspondence of coordinates 
between the SPH distributions and the horizontal axis. }\label{3models}
\end{center}
\end{figure}


\begin{thebibliography}{}
\bibitem[Abel, Bryan, Norman(2000)]{Abel00}
Abel, T., Bryan, G. L., \& Norman, M. L. 2000, \apj, 540, 39
\bibitem[Abel, Bryan, Norman(2002)]{Abel02}
Abel, T., Bryan, G. L., \& Norman, M. L. 2002, Science, 295, 93 
\bibitem[Alvarez et al.(2006)]{Alvarez06} 
Alvarez, M.~A., Bromm, V., \& Shapiro, P.~R.\ 2006, \apj, 639, 621 
\bibitem[Baraffe, Heger \& Woosely (2001)]{Baraffe01}
Baraffe, I., Heger, A. \& Woosely, S.E. \  2001, \apj, 550, 890
\bibitem[Bromm, Coppi \& Larson(1999)]{Bromm99}
Bromm, V., Coppi, P. S., \& Larson, R. B. 1999, \apj, 527, L5
\bibitem[Bromm, Coppi \& Larson(2002)]{Bromm02}
Bromm, V., Coppi, P. S., \& Larson, R. B. 2002, \apj, 564, 23
\bibitem[Cen(2003)]{Cen03} Cen, R. 2003, \apj, 591, L5
\bibitem[Ciardi Ferrara, \& White(2003)]{Cia03} 
Ciardi, B., Ferrara, A., \& White, S. D. M. 2003, \mnras, 344, L7
\bibitem[Dijkstra et al.(2004)]{Dijkstra04} 
Dijkstra, M., Haiman, Z., Rees, M.~J., \& Weinberg, D.~H.\ 2004, \apj, 601, 666 
\bibitem[Draine \& Bertoldi (1996)]{DB96}
Draine, B. T., \& Bertoldi, F. 1996, \apj, 468, 269
\bibitem[Fuller \& Couchman(2000)]{Fuller00} 
Fuller, T.~M., \& Couchman, H.~M.~P.\ 2000, \apj, 544, 6 
\bibitem[Galli \& Palla (1998)]{GP98} 
Galli D. \& Palla F. 1998, \aap, 335, 403 
\bibitem[Glover \& Brand(2001)]{GB01} 
Glover, S.~C.~O., \& Brand, P.~W.~J.~L.\ 2001, \mnras, 321, 385 
\bibitem[Haiman et al.(1997)]{Haiman97} 
Haiman, Z., Rees, M.~J., \& Loeb, A.\ 1997, \apj, 476, 458 
\bibitem[Haiman et al.(2000)]{Haiman00} 
Haiman, Z., Abel, T., \& Rees, M.~J.\ 2000, \apj, 534, 11 
\bibitem[Kahn(1954)]{Kahn54} 
Kahn, F.~D.\ 1954, \bain, 12, 187
\bibitem[Kang \& Shapiro (1992)]{KS92}
Kang, H., \& Shapiro, P., \apj, 386, 432
\bibitem[Kitayama et al.(2001)]{Kitayama01} 
Kitayama, T., Susa, H., Umemura, M., \& Ikeuchi, S.\ 2001, \mnras, 326, 1353 
\bibitem[Kitayama et al. (2004)]{Kitayama04} 
Kitayama,T., Yoshida, N., Susa, H. \& Umemura, M., 2004, \apj, 613, 631
\bibitem[Machacek et al.(2001)]{Machacek01} 
Machacek, M.~E., Bryan, G.~L., \& Abel, T.\ 2001, \apj, 548, 509 
\bibitem[Murakami et al.(2005)]{Murakami05} 
Murakami, T., Yonetoku, D., Umemura, M., Matsubayashi, T., 
\& Yamazaki, R.\ 2005, \apjl, 625, L13 
\bibitem[Nakamura \& Umemura(1999)]{NU99} 
         Nakamura F. \& Umemura M. 1999, \apj, 515, 239
\bibitem[Nakamura \& Umemura(2001)]{NU01}
	Nakamura, F., \& Umemura, M. 2001, \apj, 548, 19
\bibitem[Oh \& Haiman (2002)]{OhH02}
Oh, P \& Haiman, Z. \ 2002, \apj, 569, 558
\bibitem[Omukai \& Nishi (1999)]{ON99}
Omukai, K. \& Nishi, R. \ 1999, \apj, 518, 64
\bibitem[Ricotti, Gnedin, \& Shull(2001)]{Ricotti01}
Ricotti, M. Gnedin, N.~Y., Shull, M. \ 2001, \apj, 560, 580
\bibitem[Ricotti \& Ostriker(2004)]{Ricotti04} 
Ricotti, M.~\& Ostriker, J.~P.\ 2004, \mnras, 350, 539
\bibitem[Scannapieco, Ferrara, \& Madau(2002)]{Scan02} 
Scannapieco, E., Ferrara, A., \& Madau, P.\ 2002, \apj, 574, 590 
\bibitem[Shapiro \& Kang (1987)]{SK87}
Shapiro, P.R., \& Kang, H., 1987, \apj, 318, 32
\bibitem[Sokasian et al.(2004)]{Sokasian04} 
Sokasian, A., Yoshida, N., Abel, T., Hernquist, L., \& Springel, V.\ 2004, \mnras, 350, 47 
\bibitem[Somerville \& Livio(2003)]{Somer03} 
Somerville, R.~S.~\& Livio, M.\ 2003, \apj, 593, 611 
\bibitem[Susa (2006)]{Susa06}
Susa, H. 2006, PASJ, in press (astro-ph/0601642)
\bibitem[Susa et al.(1998)]{SUNY98}
Susa, H., Uehara, H., Nishi, R., \& Yamada, M. 1998, 
Prog. Theor. Phys., 100, 63
\bibitem[Susa \& Umemura(2004a)]{SU04a} 
Susa, H. \& Umemura, M. 2004a, \apj, 600, 1
\bibitem[Susa \& Umemura(2004b)]{SU04b} 
Susa, H. \& Umemura, M. 2004b, \apj, 610, 5L
\bibitem[Tajiri \& Umemura(1998)]{TU98} 
Tajiri, Y., \& Umemura, M.\ 1998, \apj, 502, 59 
\bibitem[Tegmark et al.(1997)]{Tegmark97} Tegmark, M., Silk, J., 
Rees, M.~J., Blanchard, A., Abel, T., \& Palla, F.\ 1997, \apj, 474, 1 
\bibitem[Wyithe \& Loeb(2004)]{Wyi04} 
Wyithe, J.~S.~B.,~\& Loeb, A.\ 2004, \nat, 427, 815 
\bibitem[Yoshida et al. (2003)]{Yoshida03}
Yoshida, N., Abel, T., Hernquist, L. \& Sugiyama, N., 2003, \apj, 592, 645
\bibitem[Yoshida(2006)]{Yoshida06} 
Yoshida, N.\ 2006, New Astronomy Review, 50, 19 
\end{thebibliography}
\end{document}